\documentclass[11pt]{article}
\usepackage{moriond,epsfig}
\usepackage[latin1]{inputenc}
\usepackage[OT1]{fontenc}

\def\be{\begin{equation}}
\def\ee{\end{equation}}
\def\bea{\begin{eqnarray}}
\def\eea{\end{eqnarray}}

\def\Mpl{M_\mathrm{Pl}}
\def\Ve{V_\mathrm{eff}}

\def\be{\beta_\mathrm{eff}}
\def\bmat{\beta_\mathrm{m}}
\def\bg{\beta_\gamma}
\def\gsim{ \lower .75ex \hbox{$\sim$} \llap{\raise .27ex
\hbox{$>$}} }
\def\lsim{ \lower .75ex \hbox{$\sim$} \llap{\raise .27ex
\hbox{$<$}} }

\begin{document}
\title{Testing Chameleon Models in the Laboratory}

\author{ A. Weltman }

\address{Department of Applied Mathematics and Theoretical Physics, \\Cambridge University, Cambridge CB2 0WA, United Kingdom. \\ Department of Mathematics and Applied Mathematics,\\ University of Cape Town, Private Bag, Rondebosch, South Africa, 7700 }

\maketitle\abstracts{
We review some recent developments in chameleon models. In particular we discuss the possibility of chameleons coupling both to photons and baryonic matter with different coupling strengths. We will discuss the possibility of probing the chameleon-photon coupling with quantum vacuum experiments in the laboratory.}

\section{Introduction}

Perhaps the greatest surprise of modern cosmology was the observation that the universe is not only expanding but is accelerating in its expansion. While the current data are consistent with the expansion being driven by a cosmological constant, dark energy is more generally modeled by  a scalar field rolling down an almost flat potential. In order for such a scalar field to evolve cosmologically today, its mass must be of order $H_0$, the present Hubble parameter. Thus, one would naively expect such a field to be essentially massless on solar system scales. A natural question then arises: if such a nearly massless field exists, why have we not detected it in local tests of the Equivalence Principle (EP) and fifth force searches? Fifth force searches put stringent bounds on the gravitational couplings allowed by light fields; bounds that would be considered unnatural from a theoretical standpoint. \\

\noindent
In this article, we will discuss a novel solution to this problem called chameleon models or the chameleon effect, whereby the coupling of a light scalar field to matter is effectively suppressed via a background dependent induced effective mass for these fields. In particular we will look at the possibility that these fields may be observed in quantum vacuum experiments in the laboratory. 

\section{Chameleon Theories}

In recent years, chameleon fields were introduced \cite{chamKW} to relieve the above mentioned tension between theory and observation. These fields allow for interesting cosmological evolution of light scalars while evading all experimental tests of gravity so far \cite{chamKW,chamcos}. Most enticingly, they offer a dark energy candidate whose properties can be probed in our local environment; via local tests of gravity in space and tests of quantum field theory in the lab. The essential feature of these fields is that due to their couplings to matter they acquire an effective potential and thus an effective mass that depends quite sensitively on the background energy density. In this way they have managed to evade all tests of the equivalence principle and fifth force experiments so far \cite{chamKW,Upadhye}.  This is not the case for space based tests of gravity as the far lower density of space allows for non-trivial and exciting predictions from chameleon models with gravitational strength coupling. On cosmological scales, the background energy density is far lower and dependent on the details of the model, these fields are a candidate for Dark Energy as well \cite{chamcos}. The choice of the name chameleon should now be clear. Not only can these fields use their local environment to hide from our tests but their properties change dependent on their environment. For extensive details on chameleon physics see \cite{chamKW,chamcos,chamstrong}.

\subsection{Ingredients}

Chameleon fields coupled to matter and photons have an action of the form 
 \begin{equation}
S = \int d^4x \sqrt{-g} \left(\frac{1}{2M_{pl}^2} R - \partial_\mu \phi \partial^\mu \phi - V(\phi)\right) 
  - \frac{e^{\phi/M_{\gamma}}}{4}F^{\mu\nu}F_{\mu\nu} + S_m(e^{2\phi/M^i_m} g_{\mu \nu}, \psi^i_m) \label{e:action} \,\, ,
\end{equation}
where $S_m$ is the action for matter and in general $\phi$ can couple differently to different matter types $\psi_i$, and $V(\phi)$ is the chameleon self interaction.  For simplicity here we will consider a universal coupling to matter defined by $\beta_m = \Mpl/M_m$. We allow for a different coupling to electromagnetism, $\beta_{\gamma} = \Mpl/M_{\gamma}$, through the electromagnetic field strength tensor $F_{\mu \nu}$.  A term of this form was originally looked at in \cite{chamcos} as a way for variations in $\phi$ to induce cosmological variations in $\alpha_{\rm EM}$, the fine-structure constant. Not all chameleon theories include a coupling to photons via an $F^2$ term as above, however recently, interest in a term of this form has been rejuvenated by a flurry of very interesting papers looking at the possibility that such a term is testable in the lab via quantum vacuum experiments \cite{PVLAS,chamPVLAS,alpenglow,gies}. 

\subsection{Effective Potential}

\begin{figure}[h]
\begin{center}
\psfig{figure=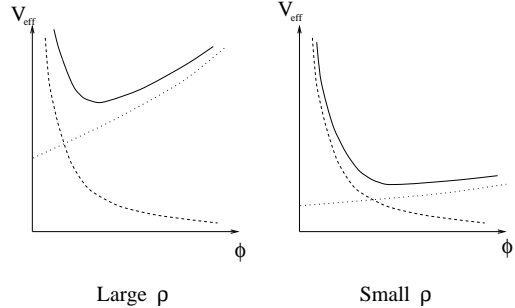,height=40mm}
\end{center}
\caption{The chameleon effective potential $V_{\rm eff}$ (solid curve) is the sum of two contributions: one from the actual potential $V(\phi)$ (dashed curve), and the other from its coupling to $\rho_m$ or $\rho_\gamma$ (dotted curve). Here, we have included the effective potential for large and small $\rho$, respectively. This illustrates that, as $\rho$ decreases, the minimum shifts to larger values of $\phi$ and the mass of small fluctuations decreases.}
\label{lrhosrho}
\end{figure}

The non-trivial coupling to matter and the electromagnetic field induces an effective potential 
\begin{equation}
\Ve(\phi,\vec x) = V(\phi) + e^{\beta_m\phi/\Mpl} \rho_m(\vec x) + e^{\beta_\gamma\phi/\Mpl} \rho_\gamma(\vec x),
\end{equation}
where we have defined the effective electromagnetic field density $\rho_\gamma = \frac{1}{2}(|\vec B^2-|\vec E|^2)$ rather than the energy density.  The presence of matter and electromagnetic fields induces a minimum $\phi_\mathrm{min}$ in $\Ve$ whereas $V$ can be a monotonic function. The dependence of this minimum on the background matter and electromagnetic fields causes the effective mass of the chameleon field to change in response to its environment. One can see this most effectively with a diagram, see Figure \ref{lrhosrho}. 
\noindent
For an exponential potential we have,
\begin{equation}
V(\phi) = \Lambda^4 \exp\left(\frac{\Lambda^n}{\phi^n}\right) \label{exp},    \quad 
\phi_\mathrm{min} \approx \left(\frac{n \Mpl \Lambda^{n+4}}{\bmat \rho_m + \bg \rho_g}\right)^{\frac{1}{n+1}} \, \, 
{\rm and} 
\,\,\, m_\phi^2 \approx  \frac{(n+1)} {\left(n \Lambda^{n+4}\right)^{\frac{1}{n+1}}} \left(\beta_m \rho_m +  \bg \rho_\gamma \right)^{\frac{n+2}{n+1}}   \\ \nonumber
\end{equation}
where the next to leading terms are suppressed by factors of $\beta_i \phi / \Mpl  \ll 1$. Thus we see explicitly here that the effective mass of the field $\phi$ is dependent on the local density of matter and  electromagnetic fields and will largely be determined by the more dense of the two. This of course also depends on the bare couplings $\beta_m$ and $\beta_\gamma$. These couplings are effectively suppressed for certain bodies, in a way to be made precise shortly. This suppression is due to the so called thin shell effect \cite{chamKW} which we will describe here. 

\subsection{Thin Shell Effect}

Consider a spherical body of homogeneous density $\rho_c$, radius $R_c$ and total mass $M_c=4\pi R_c^3\rho_c/3$, immersed in a homogeneous medium of density $\rho_\infty$. At short distances, the total force $F$, gravitational plus chameleon-mediated, on a test mass is~\cite{chamKW}
\begin{equation}
F= (1+\theta) F_N\,,
\end{equation}
where $F_N$ is Newtonian force and $\theta = 2\beta^2 $ is the fractional force due to the chameleon. For objects with large Newtonian potential, $\Phi_c= M_c/8\pi M_{Pl}^2R_c$, one finds that 
\begin{equation}
\beta \rightarrow \beta_{\rm eff} = 3 \frac{\Delta R_c}{R_c} \beta \label{theta}
\end{equation}
where $\frac{\Delta R_c}{R_c}= \frac{\phi_\infty-\phi_c}{6\beta M_{Pl}\Phi_c}$, $\phi_c$ and $\phi_\infty$ are the field values which minimize the effective potential for the respective densities. Thus, for objects satisfying the thin shell condition \cite{chamKW}, $\frac{\Delta R_c}{R_c}\ll 1$, one has $\theta\ll 2\beta^2$ and the fifth force is suppressed. Notice also that for objects displaying the thin shell effect the effective coupling, $\beta_{\rm eff}$ is independent of $\beta$. Thus strong coupling is not ruled out by experiments done to date\cite{chamstrong}. Since the thin shell effect that renders the effective coupling $\beta$ independent requires $\frac{\Delta R_c}{R_c} \ll 1$, the larger the coupling, the more likely an object will be to satisfy the thin shell condition. This new mass scale opens up the possibility for testing a large range of parameter space for chameleon fields in table top type experiments on earth. In addition, general chameleon fields may couple strongly to photons through their electromagnetic interaction. This in turn has opened a new window on testing the chameleon parameter space through experiments involving shining light through a magnetic field \cite{PVLAS,chamPVLAS,alpenglow,gies,gammevwebpage}. 

\section{Tests in the Lab : The GammeV Experiment}

In the chameleon theories discussed so far, we have used the unique properties of the chameleon to explain why they have not yet been observed in tests of gravity done on earth. We will now turn to the intriguing possibility that the very nature of these particles would make them observable in quantum vacuum experiments. Essentially in such experiments one probes the term in the action that describes the coupling of chameleons to photons via the electromagnetic interaction; 
$ \sim  \frac{e^{\phi/M_{\gamma}}}{4}F^{\mu\nu}F_{\mu\nu}$. \\

\noindent
Consider a vacuum chamber, whose walls, of density $\rho_\mathrm{wall}$, are much thicker than the chameleon Compton radius associated with the density $\rho_\mathrm{wall}$. Chameleons in the vacuum will be nearly massless and unaffected by the chameleon field in the chamber wall. As a chameleon particle approaches the wall, its mass will increase.  If its momentum is less than the chameleon mass inside the wall, then it will bounce elastically off the wall.  Thus the vacuum chamber will serve as a ``bottle'' for chameleon particles.\\

\begin{figure}[h]
\begin{center}
\psfig{figure=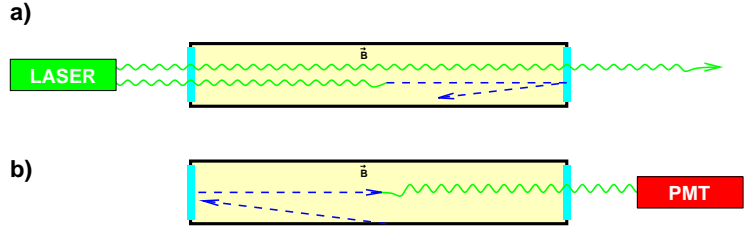,height=30mm}
\end{center}
\caption{Schematic of the GammeV apparatus. a) Chameleon production phase: photons propagating through a region of magnetic field oscillate into chameleons. Photons travel through the glass endcaps whereas chameleons see the glass as a wall and are trapped. b) Afterglow phase: chameleons in the chamber gradually decay back into photons and are detected by a photomultiplier tube.}
\label{Apparatus}
\end{figure}

\noindent
The GammeV experiment at Fermilab \cite{gammevwebpage} is just such a chameleon bottle \cite{chamgammev}.  If the chameleon field couples to photons as well as to baryonic matter, then a photon interacting with a magnetic field in the vacuum chamber will oscillate into a chameleon particle.  When this superposition of photon and chameleon states hits a glass window on one side of the chameleon bottle, it will be measured in the quantum mechanical sense; photons pass through the glass window, while chameleons are reflected.  A continuous source of photons entering the bottle will gradually fill it with a gas of chameleon particles.  After the photon source is turned off, chameleons will decay back into photons, which can escape the bottle through the glass window.  GammeV looks for this ``afterglow'' effect, a unique signature of photon-coupled chameleon particles. See figure \ref{Apparatus} for a schematic of the experiment apparatus. For a more detailed description of the apparatus, the experiment and the results see \cite{gammevwebpage} and forthcoming publications \cite{gammevcham}. 

\section{Conclusions}
 
In this short article we have reviewed the intriguing possibility that a scalar field may couple directly to both matter and photons. The rich phenomenology of these chameleon models makes concrete predictions for tests in space as well as in the laboratory possible providing us with complementary ways of testing the theory by probing different regions of parameter space. Most excitingly, it is a very real possibility that within the coming decade chameleon fields could either be observed or ruled out entirely using the space tests, laboratory tests and astrophysical observations becoming available to us. 

\section*{Acknowledgments}
A.W would like to thank the Organisers and the participants of the 43rd Recontres de Moriond conference for an excellent and productive meeting. A.W. would like to thank the GammeV collaboration and especially Amol Upadhye and Aaron Chou for many useful discussions and for their insights onto chameleon phenomenology.


\section*{References}
\begin{thebibliography}{99}

\bibitem{chamKW}  J.~Khoury and A.~Weltman, Phys.\ Rev.\ Lett.\  {\bf 93}, 171104 (2004); Phys. Rev. D {\bf 69}, 044026 (2004).

\bibitem{chamcos} Ph.~Brax, C.~van de Bruck, A.-C.~Davis,
J.~Khoury and A.~Weltman, Phys. Rev. D {\bf 70}, 123518 (2004).

\bibitem{Upadhye} A. Upadhye and S.~S.~Gubser and J. Khoury, Phys.\ Rev. \ D {\bf 74} (2006). 

\bibitem{PVLAS} E. Zavattini \emph{et al.} [PVLAS Coll.], Phys. Rev. Lett. {\bf 96}, 110406 (2006).

\bibitem{chamPVLAS} Ph.~Brax, C.~van~de Bruck, A.~C.~Davis, {\tt arXiv:hep-ph/0703243}

\bibitem{alpenglow} M.Ahlers, A. Lindner, A. Ringwald, L. Schremp and C. Weninger, {\tt arXiv:0710.1555}

\bibitem{gies} H. Gies, D.F. Mota and D. J, Shaw, {\tt arXiv:0710.1556}

\bibitem{chamstrong} D. F. Mota and D. J. Shaw, Phys. Rev. D.\textbf{\ 75}, 063501  (2007); Phys.\ Rev.\ Lett.\  {\bf 97} (2006) 151102.

\bibitem{gammevwebpage} http://gammev.fnal.gov/

\bibitem{chamgammev} The chameleon experiment described here was developed with Aaron Chou, Jason Steffen, Amol Upadhye,  and William Wester, working closely with the GammeV collaboration.

\bibitem{gammevcham} The GammeV Collaboration - in preparation. 

\end{thebibliography}
\end{document}